\documentclass[lettersize,journal]{IEEEtran}
\usepackage{amsmath,amsfonts}
\usepackage{array}
\usepackage[caption=false,font=normalsize,labelfont=sf,textfont=sf]{subfig}
\usepackage{textcomp}
\usepackage{stfloats}
\usepackage{url}
\usepackage{verbatim}
\usepackage{graphicx}
\usepackage{cite}

\usepackage[hidelinks]{hyperref}
\hypersetup{
  colorlinks = true,
  citecolor=blue,
  anchorcolor=blue,
  linkcolor=blue}

\usepackage{multirow}
\usepackage{makecell}

\usepackage[ruled,norelsize,vlined,linesnumbered]{algorithm2e}
\usepackage{soul}
\soulregister\cite7
\soulregister\ref7

\makeatletter
\newcommand{\removelatexerror}{\let\@latex@error\@gobble}
\makeatother

\begin{document}

\title{Online Mean Estimation for Multi-frame Optical Fiber Signals On Highways}

\author{Linlin Wang, Mingxue Quan, Wei Wang, Dezhao Wang, and Shanwen Wang
\thanks{The corresponding author for this article is Shanwen Wang.}
        
\thanks{Linlin Wang, Mingxue Quan, Wei Wang, and Shanwen Wang are with the School of Mathematics, Renmin University of China, Beijing 100872, China. (e-mail: wanglinlin-@ruc.edu.cn; quanmingxue@ruc.edu.cn; wwei@ruc.edu.cn; s\_wang@ruc.edu.cn)}

\thanks{Dezhao Wang is with Beijing Jhbf Technology Development Co., Ltd., China. (e-mail: 57547730@qq.com)}

}

\markboth{}%
{Shell \MakeLowercase{\textit{et al.}}: A Sample Article Using IEEEtran.cls for IEEE Journals}

\IEEEpubid{}

\maketitle

\begin{abstract}

In the era of Big Data, prompt analysis and processing of data sets is critical. Meanwhile, statistical methods provide key tools and techniques to extract valuable insights and knowledge from complex data sets. This paper creatively applies statistical methods to the field of traffic, particularly focusing on the preprocessing of multi-frame signals obtained by optical fiber-based Distributed Acoustic Sensing (DAS) system. An online non-parametric regression model based on Local Polynomial Regression (LPR) and Variable Bandwidth Selection (VBS) is employed to dynamically update the estimation of mean function as signals flow in. This mean estimation method can derive average information of multi-frame fiber signals, thus providing the basis for the subsequent vehicle trajectory extraction algorithms. To further evaluate the effectiveness of the proposed method, comparison experiments were conducted under real highway scenarios, showing that our approach not only deals with multi-frame signals more accurately than the classical filter-based Kalman and Wavelet methods, but also meets the needs better under the condition of saving memory and rapid responses. It provides a new reliable means for signal preprocessing which can be integrated with other existing methods.

\end{abstract}

\section{Introduction}

\IEEEPARstart{A}{ccurate} real-time traffic sensing is of great significance in Intelligent Traffic Systems (ITS). 
Distributed Acoustic Sensing (DAS) system is a relatively recent development for the measurement of vibration monitoring and has been used in the field of traffic\cite{daley2013field}. It uses fiber-optic cables installed next to the road for data and communication networks (telephone, internet), as a distributed detector\cite{litzenberger2021seamless}. DAS systems allow the whole time and space traffic state perception with advantages of anti-electromagnetic interference, low long-distance transmission loss, convenient installation, good concealment, and corrosion resistance\cite{wang2019comprehensive}. There have been many scholars applying DAS systems to the field of transportation, such as traffic flow detection\cite{liu2018traffic}, vehicle detection and classification\cite{liu2019vehicle}, vehicle trajectory extraction\cite{wang2021vehicle} and vehicle speed estimation\cite{wiesmeyr2021distributed}. In the application of DAS systems, there are mainly two steps for signal processing. Firstly, original signals need to be preprocessed to improve the resolution; Secondly, algorithms need to be designed to detect and track vehicles accurately and automatically from denoised DAS recordings. 

\begin{figure}[htbp!]
    \centering
    \subfloat[Lambda Architecture]{
        \includegraphics[width=3.5in]{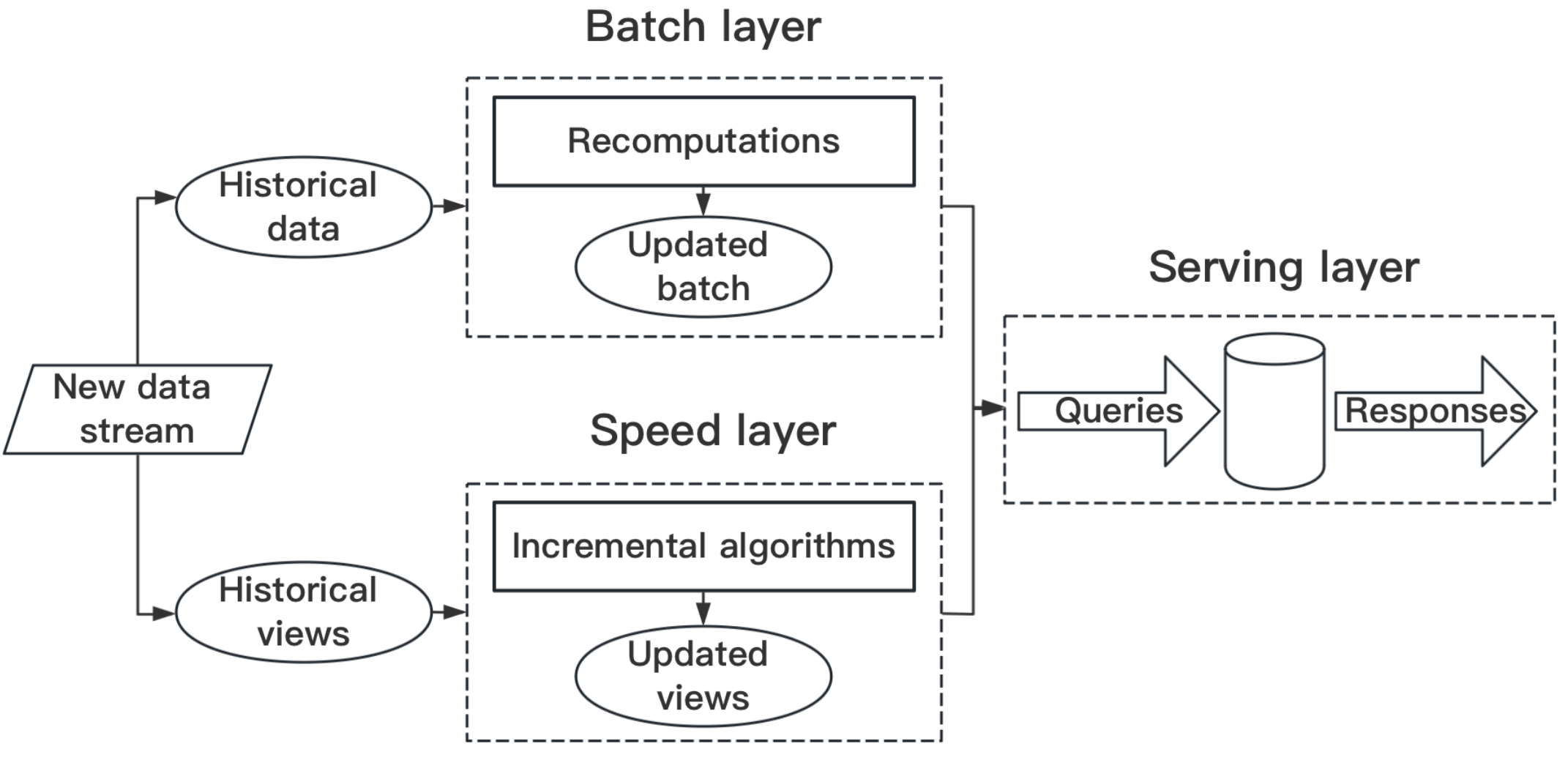}
    }\hfill
    \subfloat[Online Architecture]{
        \includegraphics[width=3.5in]{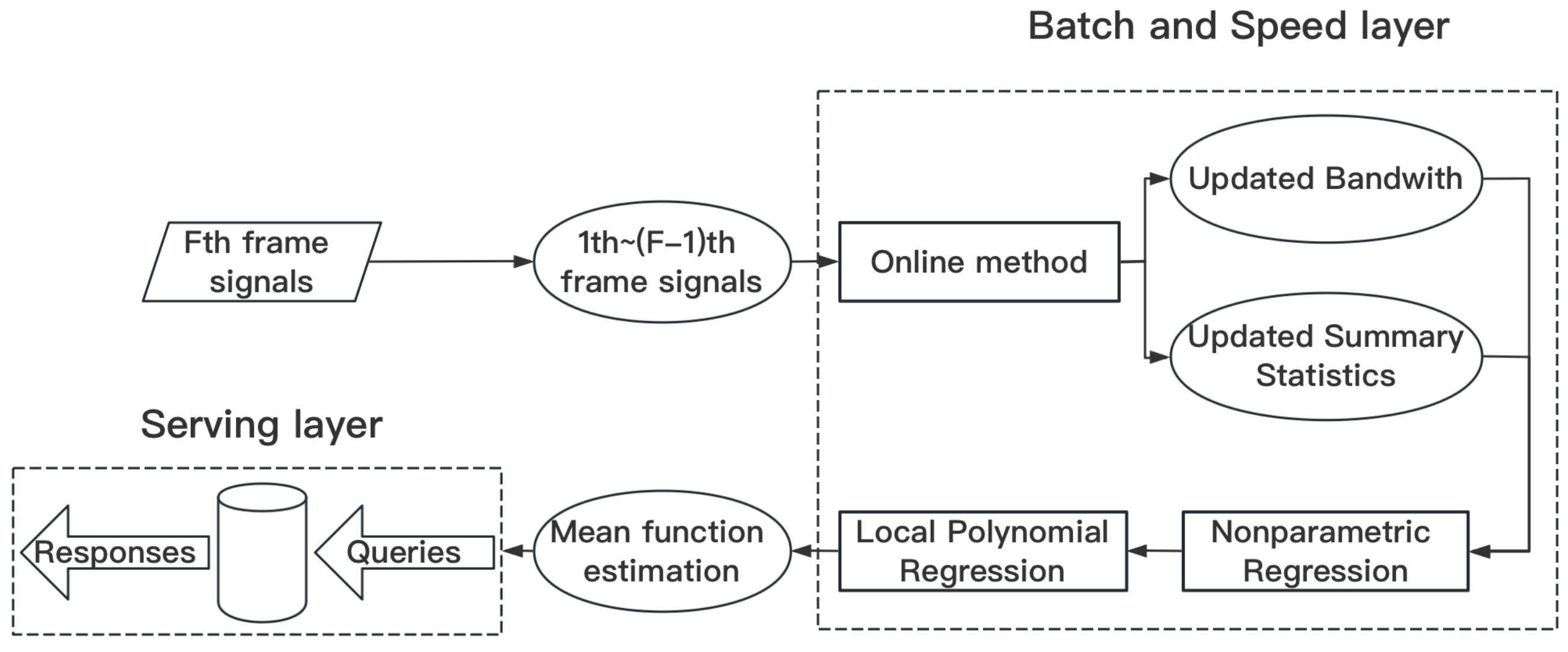}
    }\hfill
    \caption{Lambda architecture concerning new data stream flows. (a) Basic lambda architecture, and (b) online architecture in this paper. The goal of incremental algorithms utilized here is to update previously stored statistics with an incoming batch of data sets.}
\label{online_lambda}
\end{figure}

In the preprocessing step of optical fiber signals, in addition to traditional filter-based wavelet denoising\cite{liu2018traffic}, many scholars pay attention to artificial intelligence algorithms. A convolutional neural network (CNN) algorithm based on an attention mechanism with an extremely short signal window to denoise signals is proposed in\cite{wang2021rapid}. A self-supervised time-domain deconvolution Auto-Encoder (time-domain DAE) is proposed in\cite{van2022deep}, which uses a stationary Ricker wavelet in the time domain as the vehicle's impulse response to convolve signals along the time axis. A developed spatial-domain deconvolution Auto-Encoder (spatial-domain DAE) uses kernel described by the Flamant-Boussinesq proximation\cite{yuan2020near},\cite{jousset2018dynamic} to convolve signals along the space axis and proves its effectiveness\cite{yuan2023spatial}. However, these deep learning algorithms require a large amount of signals to train, and the models are poorly interpretable.

In the detection and tracking step of preprocessed optical fiber signals, most scholars convert DAS signals into two-dimensional waterfall diagrams and regard vehicle trajectory as a straight line and extract its peaks to derive velocity. Through the conversion of the polar coordinate system, the line detection methods Radon transform\cite{wang2021vehicle} and Hough transform\cite{wiesmeyr2021distributed} are used to obtain the singularity of the range-velocity domain. K-means clustering and Kalman filtering are used to determine the locations of individual vehicles and track the trajectories of vehicles over time\cite{thulasiramantraffic}. The double threshold algorithm based on the combination of short-term energy and short-term zero-crossing rate is used for detection\cite{liu2019vehicle}. However, none of these methods can capture instantaneous velocity and realize real-time tracking. In addition, it is difficult to distinguish the trajectories of different vehicles in an environment of traffic jams.

The demand for efficient communication and data storage in traffic monitoring is continuously increasing, signal representation and compression are quite important factors\cite{engan2000frame}. Particularly in the field of optical fiber application, vehicle signals collected by DAS systems are enormous and multi-frame, causing trouble in data redundancy, data asynchronism, processing complexity, memory occupation and real-time requirements\cite{pokorny2015big}. Thus signals need to be preprocessed and compressed according to demands. Traditional signal decompositions such as filter-based Wavelet and Kalman methods, can reconstruct signals from limited observation sets, but are not able to perform data compression. The ineffectiveness of common tools in the field of transportation has led us to explore new ways in the statistical sense. 

The signals collected by DAS systems can be viewed as streaming data from the statistical perspective, which refers to a set of data sequences coming in continuously and rapidly, but there is limited memory to store them\cite{muthukrishnan2005data}. Furthermore, signals are recorded at several discrete distance points and have similar properties with functional data, which originates from continuous functions, but due to practical limitations such as technological constraints, are sampled discretely\cite{wang2016functional}. In terms of definition and properties, multi-frame problems can be abstracted into online mean estimation problems of functional data. Moreover, in statistical thought, the trajectory of each subject can be viewed as a stochastic process, while the mean function is modeled non-parametrically with information accumulated to improve the accuracy of estimation\cite{ferraty2006nonparametric}. In addition, optical fiber signals have spatial and temporal characteristics, and a series of exploratory methods for spatio-temporal data is applicable in the field of statistics, such as visualization, spectral analysis, Empirical Orthogonal Function (EOF) analysis, Canonical Correlation Analysis (CCA), etc\cite{cressie2015statistics}. In the above regard, this paper attempts to employ statistical methods regarding functional data to process optical fiber signals.

Since optical fiber signals are dynamically flowing frame by frame during the acquisition process, signals are required to be processed and analyzed immediately as soon as they enter the system, which is concluded as online processing (also known as real-time processing). With the emerging field of Big Data analytics and real-time applications, numerous online processing methods and technologies have been developed and applied\cite{zheng2019real}. Among them, statistical online algorithms can better improve the timeliness of information processing, and can also update the database online, which can better connect the database and provide technical support for data preprocessing\cite{hammer2017big}. Online statistical methods put more emphasis on online processing, performing detection or classification based on information from the past and present, without seeing the future. Based on the non-parametric model and regression method, this research can update bandwidth and summary statistics as signals flow rapidly, boasting three major advantages: (1) It is based on a statistical method and has strong interpretability; (2) it reflects the overall trend and can be updated with a rapid response; and (3) using only summary statistics of previous data, instead of storing entire data is of critical importance in saving computing memory and speed. This online architecture for streaming data analysis can be implemented in the Lambda Architecture design pattern currently used by Twitter and AWS\cite{warren2015big}. Fig. \ref{online_lambda}(a) shows the standard Lambda Architecture, and Fig. \ref{online_lambda}(b) shows the specific application here.


This paper is devoted to dealing with multi-frame problems of optical fiber signal preprocessing via statistical methods in real time. An online mean estimation approach based on non-parametric regression is employed\cite{fang2021online}. This approach utilizes Local Polynomial Regression (LPR) and Variable Bandwidth Selection (VBS) to estimate the mean value of multi-frame signals in real time under computing memory constraints. This online estimation approach does not have to store all data sets, but storing two summary statistics is sufficient to achieve great accuracy comparable to that of using the entire dataset, which are means of signal compression. The mean estimator is renewed with current streaming data sets and summary statistics of historical data, which is the process of signal reconstruction. The work of this paper is to apply this online algorithm in the realm of Intelligent Transportation Systems, enabling the estimation of mean functions based on historical summary statistics in conjunction with current data. Additionally, we go further to extract fiber stripes from waterfall diagrams to validate the impact of the algorithm on subsequent vehicle trajectory extraction algorithms. The statistical method used in this paper is compared with filter-based methods, considering 1-D signal amplitudes and 2-D waterfall diagram information, showing its effectiveness in practical highway scenarios.

The main contributions of this paper are as follows:
\begin{enumerate}
    \item We abstract signal processing problems in DAS systems as a non-parametric regression problem in the field of statistics.
    \item We use an online method to extract summary statistics for signal compression with little memory usage. As signals flow in frame by frame, the statistics can be updated in real time and signal can be reconstructed with rapid responses.
    \item Our proposed technique can not only deal with the preprocessing of multi-frame rate signals, but also be applied to time series data conforming to the same model.
\end{enumerate}

The remainder of this paper is as follows. Section \ref{related_works} introduces DAS technology, local polynomial regression and online lambda architecture. Section \ref{method} displays the methodology, including online mean estimation method and trajectory extraction algorithm. Section \ref{experiments} introduces the design of experiments and discusses the experimental result. Section \ref{conclusion} provides conclusions and limitations of this study.

\section{Related Works}
\label{related_works}

\subsection{Distributed Acoustic Sensing Technology}

Distributed Acoustic Sensing (DAS) is a rapidly developed passive fiber-optic sensing technology, which can detect acoustic signals anywhere along the length of fiber with high-frequency responses and tight spatial resolution\cite{cannon2013distributed}. DAS not only has the advantages of traditional fiber-optic sensing technologies (e.g., anti-electromagnetic interference, corrosion resistance, slenderness, and flexibility) but can also measure dynamic strains like vibrations and sound waves along fiber paths in a long-distance, fully distributed, and online manner. Given the considerable application values of DAS, a growing number of experts and scholars are exploring its possibility in Intelligent Transportation Systems (ITS). They are exploring its potential for vehicle traffic monitoring, making traffic detection methods based on the DAS system a focal point of recent research and development. Detection of vehicle speed, density, and road conditions can be achieved using fiber-carrying high-speed data transmission\cite{wellbrock2019first}. 

Generally speaking, the Optical Fiber System, highlighted by its DAS technology, boasts three major advantages. Firstly, its all-weather capability ensures effective operation under a variety of adverse weather conditions such as rain, snow, fog, and wind, maintaining functionality both during the day and night. Secondly, the system offers comprehensive tracking by enabling online monitoring of all vehicles over extensive road sections, thereby eliminating the necessity for integration with multiple poles or sensors. Lastly, it showcases green and cost-efficiency as a single fiber strand can span up to 40 kilometers, significantly cutting down on construction and maintenance costs while exemplifying green energy efficiency.

\subsection{Non-parametric Regression Model}
\label{NRM}

The model adopted in this paper is an online non-parametric regression model based on Local Polynomial Regression (LPR) and Variable Bandwidth Selection (VBS), which is always utilized for smoothing scatter plots and modeling functions\cite{avery2013literature}, so its principle and formula are summarized here.

Suppose $\{(x_{i}, y_{i})| x_{i} \in \mathbb{R}^{n}, y_{i} \in \mathbb{R}^{n}\}_{i=1}^{n}$ are random samples from a non-parametric regresssion model:
\begin{equation}
    y_i = f(x_i)+ \epsilon_i
\end{equation}
where $y_i$ is the response variable, $x_i$ is the explanatory variable, $f(\cdot)$ is a non-parametric smoothing function, and $\epsilon_{1},..., \epsilon_{n}$ are uncorrelated random errors with mean 0 and variance $\sigma^2$. Now we need to solve $E(y_i|x_i) = f(x_i)$ for its best estimation $\widehat{f}(x_i)$.

The main idea of LPR is to estimate $f(\tilde{x})$ using points $x$ adjacent to the current point $\tilde{x}$. Using the Taylor expansion and taking the first $p+1$ terms, the formula can be simplified as:
\begin{equation}
    f_p(x; \tilde{x}) = \beta_0 + \sum_{j = 1}^{p}\beta_j(x - \tilde{x})^j
\label{f(x)}
\end{equation}
where $\beta_0 = f(\tilde{x}), \beta_j = \frac{f^{(j)}(\tilde{x})}{j!}$, and $p$ is the order of polynomial fit.

LPR introduces the concept of weights and utilizes a low-order weighted least squares (WLS) regression to fit. The weight $\omega$ is usually measured by kernel function $K_h(\cdot)$, which depends on bandwidth $h$. Denote the weight $\omega_i$ is the weight between $\tilde{x}$ and its neighbor point $x_i$, and a one-dimensional kernel function is used here, then the weight can be expressed as:
\begin{equation}
    \omega_i = K_h(x_i; \tilde{x}) = \frac{1}{h}K(\frac{x_i - \tilde{x}}{h})
   \label{kernel}
\end{equation}

The least squares estimate of the model is usually obtained by minimizing the weighted Residual Sum of Squares (RSS). Through multiplying the weight function $\omega_i$ and the loss $(y_i - f_p(x_i; \tilde{x}))^2$ at each point around $\tilde{x}$, RSS can be written as:
\begin{equation}
    RSS(\beta) = \sum_{i = 1}^{n}(y_i - f_p(x_i; \tilde{x}))^2\omega_i = (y - \Phi\beta)^{T}\Omega(y - \Phi\beta)
\end{equation}
where $\Omega = {\rm {diag}}(\omega_1, \omega_2,..., \omega_n), y = (y_1, y_2,..., y_n)^{T}, \beta = (\beta_{0},\beta_{1},..., \beta_{p})^{T}$,
\begin{equation}
    \Phi = \begin{pmatrix}
         1 & x_1 - \tilde{x} & (x_1 - \tilde{x})^2 & \cdots & (x_1 - \tilde{x})^p \\
         1 & x_2 - \tilde{x} & (x_2 - \tilde{x})^2 & \cdots & (x_2 - \tilde{x})^p \\
        \vdots & \vdots & \vdots & \ddots & \vdots \\
        1 & x_n - \tilde{x} & (x_n - \tilde{x})^2 & \cdots & (x_n - \tilde{x})^p
         \end{pmatrix}
\end{equation}

Take the partial derivative of $\beta$ to minimize the loss function, and we get the estimate of $\beta$:
\begin{equation}
    \widehat{\beta} = (\Phi^T\Omega\Phi)^{-1}\Phi^T\Omega y 
    \label{beta}
\end{equation}

Plug the above estimates into the formula (\ref{f(x)}) and the mean estimation for each sample point is:
\begin{equation}
    \widehat{y}_i = \widehat{f}(x_i; \tilde{x}) = \widehat{\beta}_0 + \sum_{j = 1}^{p}\widehat{\beta}_j(x_i- \tilde{x})^j
\end{equation}

As can be seen from the above, LPR is a kernel-based estimator, which has the form of locally weighted least squares. According to the theoretical results of optimal variable bandwidths, which would be the ideal bandwidth to work with, $h \propto \frac{1}{n}$\cite{fan1995data}. Therefore, the bandwidth should be indicated by the streaming data themselves, and be updated by a data-driven VBS procedure.

\subsection{Lambda Architecture}

Lambda architecture is a Big Data system of computing and storage with synchronized processing of batch and stream data flows, consisting of three layers\cite{luo2020renewable}: 

\begin{enumerate}
    \item \textbf{Batch layer} continuously recomputes previously-stored batch views when a new batch of data arrives, achieving data accuracy by processing all existing historical data.
    \item \textbf{Speed layer} utilizes incremental algorithms to update old online views with an incoming batch of data, minimizing latency by providing a real-time view of the newest data.
    \item \textbf{Serving layer} stores the view outputs of the above two layers and responds to queries.
\end{enumerate}

Not only the data integrity is taken into account through the batch layer, but also the high delay of batch layer can be compensated by speed layer, making the whole query real-time. The basic lambda architecture of how it works is illustrated in Fig. \ref{online_lambda} (1). The lambda architecture enables us to facilitate sequential updating of summary statistics used in estimation, and is flexible and scalability to a wide range of streaming data analyses in which the batch layer stores all accumulated data and produces reliable results. Moreover, it enables developers to build large-scale distributed data processing systems and is tolerant of hardware failure and human error.

\section{Method}
\label{method}
\subsection{Problem Formulation}

In the acquisition process of the Distributed Acoustic Sensor (DAS) system, the length of optical fiber is fixed and we denote sampling points as $d \in \{1,2,...,D \}$. Denote the acquisition Frames Per Second (FPS) in optical fiber systems by $F_{\rm max}$ and the max number of seconds collected by $S$. Then signal amplitude at each frame-distance vector can be expressed as $A^s_f = (a^s_{f1}, a^s_{f2},..., a^s_{fD}), f \in \{1,2,..., F_{\rm max}\}, s \in \{1,2,..., S\}$, where $a^s_{fd} \in \mathbb{R^{+}}$ is signal amplitude. If the data matrix collected at second $s$ is denoted as $A^s$, it can be expressed as:
\begin{equation}
    A^s = \begin{pmatrix}
        A^s_1 \\
        A^s_2 \\
       \vdots \\
        A^s_{F_{\rm max}}
        \end{pmatrix} = 
        \begin{pmatrix}
         a^s_{11} & a^s_{12} & \cdots & a^s_{1D} \\
        a^s_{21} & a^s_{22} & \cdots & a^s_{2D} \\
        \vdots & \vdots & \ddots & \vdots \\
        a^s_{F_{\rm max}1} & a^s_{F_{\rm max}2} & \cdots & a^s_{F_{\rm max}D}
         \end{pmatrix}
    \label{matrix}
\end{equation}
    
It should be noted that the flow of signals can be regarded as streaming data over time, and if the total signals are accumulated to the second $S$, the complete matrix can be represented as $A = (A^1, A^2,..., A^S)^T$. Due to FPS setting of the DAS system, the fiber signals are multi-frame recorded and we need to export the specific number of frames according to the requirements.

Although signals are collected at discrete points, we can reduce them to 1-D continuous information for further modeling analysis. In addition, we generalize the above questions to take into account the possibility of changing the location of the collection point, which is, that the distance point and observations may vary due to artificial setup. The 1-D amplitude information $\{A_f(X_{fd}): X_{fd} \in \mathcal{X}\}$ can be considered as an $L^2$ stochastic process on a distance interval $\mathcal{X}$ for each frame $f \in \{1,..., F_{\rm max}\}$. In each frame $f$, we have a subject with $D$ observations at discrete distance point $X_{fd}, d \in \{1,..., D\}$, so observations can be written in discrete form:
\begin{equation}
    A_f(X_{fd}) = \mu(X_{fd}) + \Phi_f(X_{fd})
\end{equation}
where $\mu(\cdot)$ is mean function, $\Phi(\cdot)$ is stochastic part.
Attention that our observations are contaminated with noises in practical application. To describe the stochastic process more accurately, we introduce white noise and rewrite the observations:
\begin{equation}
    Y_{fd} = A_{f}(X_{fd}) + \epsilon_{fd} = \mu(X_{fd}) +\Phi_{f}(X_{fd}) + \epsilon_{fd}
\label{non-parametric}
\end{equation}
where $\epsilon_{fd}$ are independently and identically distributed(i.i.d) noises with $E(\epsilon_{fd}) = 0$ and ${\rm Var}(\epsilon_{fd}) = \sigma^2$.

FPS in optical fiber systems is set manually, where a higher sampling rate implies more data points to be collected and more signal details to be captured. However, when signals are output as structured data in practical applications, it is not necessary to use all multiple frames of data, and storing them consumes significant memory. Therefore, how to handle high frame rate signals in fiber-based optic systems poses a challenge. Moreover, distinguishing the trajectory of each vehicle and conducting traffic flow statistics is also a great challenge. How to deal with multi-frame signals in each frame and extract trajectories accurately is the primary objective of this paper.

\subsection{Online Mean Estimation}

Based on the above analysis of optical fiber signals and the non-parametric model, the Local Polynomial Regression (LPR) method is adopted to estimate the mean function. In our experiments, the common Epanechnikov kernel function for non-parametric filtering and data smoothing is employed as kernel function $K(\cdot)$ in (\ref{kernel}), and the formula is as follows:
\begin{equation}
    K(x)=\begin{cases}
    \frac{3}{4}(1-x^2),& |x| < 1;\\
    0,& |x| \geq 1.\end{cases}
\end{equation}

Since optical fiber signals can be seen as streaming data flowing in continuously, and bandwidth is inversely proportional to the amount of data\cite{fan1995data}. The online mean estimation algorithm pays particular attention to Variable Bandwidth Selection (VBS). Future optimal bandwidth is approximated by a corresponding online estimator, and a dynamic candidate bandwidth sequence is utilized to select the most suitable one. After that, historical summary statistics are combined to form the updated summary statistics and mean function estimations. 

For notations, the subscript $ \widehat{}$ and $\widetilde{}$ denote the classical batch method using full data and the online method using only the $f$th data and stored summary statistics respectively.

\subsubsection{Classical Batch Method}
Based on the non-parametric modeling of optical fiber signals in (\ref{non-parametric}), we can use the Local Polynomial Regression model in Section \ref{NRM} to derive mean function estimation. For the sake of simplicity, a first-order LPR model ($ p = 1 $) is used, the simple regression model is:
\begin{equation}
    Y_{fd} = \beta_0 + \beta_1X_{fd} + \epsilon_{fd}
\end{equation}

By minimizing the loss function based on the weighted Residual Sum of Squares (RSS), we get the estimation of $(\beta_0, \beta_1)$:
\begin{equation}
\begin{split}
    (\widehat{\beta}_0, \widehat{\beta}_1) =  \arg\min\limits_{\beta_0, \beta_1}
    \sum_{f = 1}^{F}\sum_{d = 1}^{D}\left\{Y_{fd} - \beta_0 - \beta_1(X_{fd} - x)\right\}^2 \\
    \times K_{\widehat{h}^{(F)}}(X_{fd} - x)
\end{split}
\end{equation}
where $\widehat{h}_\mu^{(F)}$ is the bandwidth using the full data up to $F (F \leq F_{\rm max})$. Let $e_2 = (1,0)^T$, $X_{fd}(x) = (1, X_{fd} - x)^T$, and the solution of classical mean estimation at block $F$: $\widehat{\mu}^{(F)}(x) = \widehat{\beta}_0$ can be explicitly rewritten as:
\begin{equation}
    \widehat{\mu}^{(F)}(x) = e_2^T\left\{\sum_{f = 1}^{F}P_f(x;\widehat{h}^{(F)}_\mu)\right\}^{-1}\left\{\sum_{f = 1}^{F}Q_f(x;\widehat{h}^{(F)}_\mu)\right\}
\end{equation}
where $\{P_f, Q_f\}$ only depends on the $f$th frame given bandwidth $\widehat{h}_\mu^{(F)}$ by
\begin{equation}
    P_f(x;\widehat{h}^{(F)}_\mu) = \sum_{d = 1}^{D}K_{\widehat{h}^{(F)}_\mu}(X_{fd} - x)X_{fd}(x)X_{fd}(x)^T
\label{P_f}
\end{equation}
\begin{equation}
    Q_f(x;\widehat{h}^{(F)}_\mu) = \sum_{d = 1}^{D}K_{\widehat{h}^{(F)}_\mu}(X_{fd} - x)X_{fd}(x)Y_{fd}
\label{Q_f}
\end{equation}

From the formulations above, such kernel-based mean estimator can be decomposed into two summary statistics $\sum_{f = 1}^{F}P_f(x;\widehat{h}^{(F)}_\mu)$ and $\sum_{f = 1}^{F}Q_f(x;\widehat{h}^{(F)}_\mu)$, which are additive on data depending on the tuning variable bandwidth $\widehat{h}^{(F)}_\mu$. Given the bandwidth, we only need to store a pair of precise summary statistics in computing memory instead of the entire $F$ data frames, which is highlighted in the era of Big Data. 

\subsubsection{Online Estimation}
According to the theoretical result of LPR, bandwidth is inversely proportional to the sample size. As frame $F$ varies, the sample size of signals gradually varies, bandwidth of classical batch method $\widehat{h}^{(F)}_\mu$ may take different values. As the number of frames increases to the maximum $F_{\rm max}$, this data-driven feature has to store all summary statistics $\left\{\sum_{f = 1}^{F}P_f(x;\widehat{h}^{(F)}_\mu), \sum_{f = 1}^{F}Q_f(x;\widehat{h}^{(F)}_\mu)\right\}_{F = 1}^{F_{\rm max}}$. To overcome this obstacle and respond to streaming data in real time, a method to update statistics based on dynamic candidate bandwidth sequence is adopted.

Denote $h_{\mu,*}^{(f)}$ as the optimal bandwidth of the $f$th frame for $f \le F_{\rm max}$ and $\widetilde{h}^{(f)}_\mu$ as its corresponding online estimator. The expression of both are shown in\cite[Section 4]{fang2021online}. According to\cite[Theorem 3]{fang2021online}, the online bandwidth estimator $\widetilde{h}^{(f)}_\mu$ satisfies that as $f \rightarrow \infty$:
\begin{equation}
    \frac{\widetilde{h}^{(f)}_\mu - h_{\mu,*}^{(f)}}{h_{\mu,*}^{(f)}} = O_p(f^{-\frac{1}{5}})
\end{equation}
where $O_p$ means bounded in probability\cite{awvd1998asymptotic}. The formula above implies the online estimator $\widetilde{h}^{(F)}_\mu$ converges to the optimal bandwidth $h_{\mu,*}^{(F)}$, and can serve as a surrogate of it.

According to the characteristics of streaming data and the inverse relationship between bandwidth and sample size, if the whole data is not used for calculation when new data flows in, we need to update the previous bandwidth. Following the strategies in\cite{fang2021online}, the calculation of the $F$th frame signals mean estimation is by recursion as follows.

Let $\{\eta_{\mu,l}^{(f)}\}_{l=1}^{L}$ for $f < F$ be the dynamic candidate bandwidth sequence with
\begin{equation}
    \eta_{\mu,l}^{(f)} = (\frac{L-l+1}{L})^{\frac{1}{5}}\widetilde{h}_{\mu}^{(f)}
\label{eta}
\end{equation}
and let $\{\phi_{\mu,l}^{(f - 1)}\}_{l=1}^{L}$ be the centroids (i.e.weighted average of all previous candidate bandwidths):
\begin{equation}
    \phi_{\mu,l}^{0} = 0, \phi_{\mu,l}^{(f)} = (1 - \frac{1}{f})\phi_{\mu, j_l^{(f)}}^{(f-1)} + \frac{1}{f}\eta_{\mu, l}^{(f)},
\end{equation}
where
\begin{equation}
j_l^{(f)} = \arg\min\limits_{i \in \{1,2,...,L\}}\left|\eta_{\mu, l}^{(f)} - \phi_{\mu,i}^{(f-1)}\right|.
\end{equation} 
The choice of $j_l^{(f)}$ in the formula guarantees that $\phi_{\mu,j_l^{(f)}}^{(f-1)}$ is close to $\eta_{\mu,l}^{(f)}$, which makes $\phi_{\mu,l}^{(f)}$ close to $\eta_{\mu, l}^{(f)}$ as well. As a result, we get the most suitable bandwidth $\eta_{\mu,j_l^{(f)}}^{(f)}$ and output the $f$th summary statistics whenever needed. 

To combine information across frames, the so-called “pseudo-summary statistics” $\{\widetilde{P}_{\mu,l}^{(f)}, \widetilde{Q}_{\mu,l}^{(f)}\}_{l=1}^{L}$, whose update depends only on the sub-statistics of the current frame $\{P_f, Q_f\}$ which are calculated following formulas (\ref{P_f}) and (\ref{Q_f}), and the stored pseudo-summary statistics of previous frames $\{\widetilde{P}^{(f-1)}_{\mu,j_l^{(f)}}, \widetilde{Q}^{(f-1)}_{\mu,j_l^{(f)}}\}$. 
the pseudo-summary statistics can be expressed as:
\begin{equation}
    \widetilde{P}_{\mu,l}^{(f)}(x) = P_f(x;\eta_{\mu,l}^{(f)}) + \widetilde{P}^{(f-1)}_{\mu,j_l^{(f)}}(x)
\end{equation}
\begin{equation}
    \widetilde{Q}_{\mu,l}^{(f)}(x) = Q_f(x;\eta_{\mu,l}^{(f)}) + \widetilde{Q}^{(f-1)}_{\mu,j_l^{(f)}}(x)
\end{equation}
with initialization $\widetilde{P}_{\mu,l}^{(0)}(x) = 0, \widetilde{Q}_{\mu,l}^{(0)}(x) = 0$ for $l \in \{1, 2, ..., L\}$.

It's worth noting that $\{\eta_{\mu,l}^{(f)}\}_{l=1}^{L}$ and $\{j_l^{(f)}\}_{l = 1}^{L}$ are recalculated at each $f$, and only the newest $\{\phi_{\mu,l}^{(f)}\}_{l=1}^{L}$ and $\{\widetilde{P}_{\mu,l}^{(f)}, \widetilde{Q}_{\mu,l}^{(f)}\}_{l=1}^{L}$ are stored in memory during the procedure. Hence the algorithm is computationally efficient.

Follow the recursive process above and note that $\eta_{\mu,1}^{(F)} = \widetilde{h}^{(F)} _\mu$ in formula (\ref{eta}), the mean estimation at frame $F$ is given by:
\begin{equation}
    \widetilde{\mu}^{(F)}(x) = e_2^T\left\{\widetilde{P}_{\mu,1}^{(F)}(x)\right\}^{-1}\left\{\widetilde{Q}_{\mu,1}^{(F)}(x)\right\}
\end{equation}

Hence the online mean estimation of multi-frame signals are effectively achieved.

\subsection{Trajectory Extraction}

During the trajectory extraction processing of optical fiber signals, we represent them as three-dimensional waterfall diagrams. Based on these diagrams, we utilize a vehicle position detection algorithm and a vehicle trajectory extraction algorithm to achieve detection and overall tracking for vehicles\cite{2024arXiv240209422W}. 

After the preprocessing of optical fiber signals, we use the first column information, which is processed by the peaks location search method, to determine the vehicles' entry times. Then, an algorithm for line-by-line matching based on the vehicle motion model to track key points of vehicle trajectories. Then we can reconstruct the complete vehicle trajectories and subsequently calculate the instantaneous speed of the vehicles. 

\subsubsection{Vehicle Position Detection}
First, we extract the peaks of the first column signals of waterfall diagrams by using the peaks location search method where the first derivative is zero and the second derivative is negative. Given a certain threshold, we calculate the local maximum points and retain only the peak values that are higher than the threshold as the entry time points of the vehicle.

\subsubsection{Vehicle trajectory extraction}
Based on Newton's first theorem, over a short period of time, a vehicle can be regarded as traveling at a constant velocity. In this paper, the time interval of the signal is regarded as the minimum unit, and the velocity of the vehicle is assumed to remain constant in each minimum unit.

\begin{figure}[t!]
    \centering
    \includegraphics[width=2.5in]{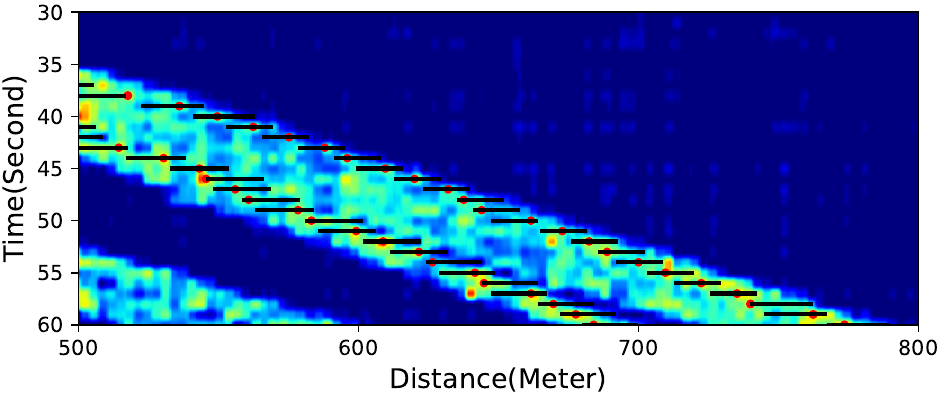}
\caption{Trajectory extraction algorithm. The black line segments represent the distance confidence interval and the red dots represent trajectory key points.}
\label{tunnel}
\end{figure}

The vehicle position detection algorithm flow is shown in Algorithm \ref{algorithm_1}. We examined the algorithm using real signals collected by the DAS system in part of the tunnel sections on highways, as shown in Fig. \ref{tunnel}. The highlighted signals show the trajectory of vehicles, while the dark blue signals show the background noise. Among them, the first column of red dots marks the collected position information at the moment of vehicle entry, the black lines are combined with the speed limit information of the tunnel section to give a certain degree of confidence, and the following red dots mark the position information of the vehicle every second. It can be found that even if the two vehicles are very close, the algorithm can extract the real-time position information of the vehicle more accurately, and then calculate the average velocity and instantaneous velocity.

\begin{algorithm}[htbp]
    \caption{Trajectory Extraction Algorithm.}
    \label{algorithm_1}
    \LinesNumbered 
    \KwIn{$D = (d_{ij})$: preprocessed optical fiber signal matrix;\ $i = 1, ..., m$, $j = 1, ..., n$, $d_{ij} \in \mathbb{R^{+}}$: selected time and distance.}
  
    \KwOut{$\Sigma^{(S)} = \{(\sigma^{(s)})_{1 \leq s \leq S}\}$: vehicle trajectory point sets.}

    $\{k_s \in \mathbb{N^{+}}, s = 1, ...,S\}$ $\gets$ FindPeak($D_{1: m,1}$), where $S$ is vehicle counts and $k_s$ is the appearance time of each vehicle\;
    $l_{k_s + \alpha}\in \mathbb{N^{+}}$ is distance position point corresponding to time position point $k_s + \alpha$ in $D = (d_{ij})$\;
    $(k_s + \alpha, l_{k_s + \alpha})$ is a pair of time and distance position points of each extracted vehicle trajectory\;

    $\Sigma^{(S)} = \{(\sigma^{(s)})_{1 \leq s \leq S}: \sigma^{(s)} = (k_s + \alpha, l_{k_s + \alpha}), \alpha \in \mathbb{N}$\}\;
    
    Select the initial velocity interval $v_{\min}^{(1)}$ to $v_{\max}^{(1)}$\;
    Calculate the initial distance interval  $x_{\min}^{(1)}$ to $x_{\max}^{(1)}$\;

    \ForEach{s = 1 to S; $\sigma^{(s)} \in \Sigma^{(S)}$}{
      Add the first pair $(k_s, l_{k_s})$ to vehicle trajectory point sets $\sigma^{(s)}$, where $l_{k_s} = 1$\;
      $l_{k_s + 1}$ $\gets$ ColIndex(Max ($D_{k_s + 1, l_{k_s}+ x_{\min}^{(1)}: l_{k_s} + x_{\max}^{(1)}}$))\;
       \eIf{$k_s + 1 < m$ $\rm and$ $l_{k_s + 1} < n$}{
          Add pair $(k_s + 1, l_{k_s + 1})$ to each vehicle trajectory point set $\sigma^{(s)}$\;
       }{
          break;
       }
      } 

    \ForEach{s = 1 to S; $\sigma^{(s)} \in \Sigma^{(S)}$}{
      \For{$\alpha \in \mathbb{N^{+}}$}{
        $y^{(s)}$ = PolynomialFitting($\sigma^{(s)}$)\;
        $v^{(s)}$ = Slope($y^{(s)}$)\;
        Select the right level of confidence $\rm cof$\;
        Calculate velocity interval $(1 + \rm cof)$$v^{(s)}$ to $(1 + \rm cof)$$v^{(s)}$\;
        Calculate distance interval $x_{\min}^{(s)}$ to $x_{\max}^{(s)}$\;
        $l_{k_s + \alpha}$ $\gets$ ColIndex(Max ($D_{k_s + \alpha, l_{k_s + \alpha - 1}+ x_{\min}^{(s)}: l_{k_s + \alpha - 1} + x_{\max}^{(s)}}$))\;
       \eIf{$k_s + \alpha < m$ $\rm and$ $l_{k_s + \alpha} < n$}{
          Add pair $(k_s + \alpha, l_{k_s + \alpha})$ to each vehicle trajectory point set $\sigma^{(s)}$\;
       }{
          break;
       }
  
      } 

      }
    \textbf{Return:} Vehicle trajectory point sets $\Sigma^{(S)}$.
\end{algorithm}

\section{Experiments}
\label{experiments}
\subsection{Datasets}

\begin{figure}[!t]
    \centering
    \includegraphics[width=3in]{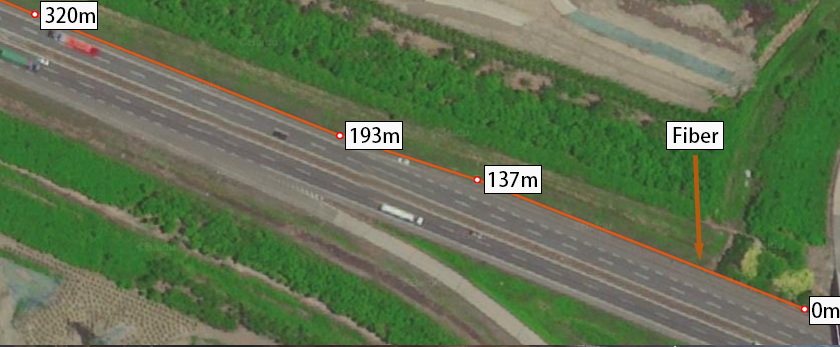}
    \caption{Overhead view of the Jingjintang expressway, and the optical fiber is laid on one side of the road.}
    \label{view}
\end{figure}

Experiments are implemented on real highway scenarios for data acquisition and data analysis. Jingjintang Expressway (Junliang City Test Site, Dongli District, Tianjin, China) is a real highway scenario of a dual carriageway with three lanes in each direction. We selected 320 meters section of Jingjintang Expressway. The aerial view of this highway is illustrated in Fig. \ref{view}.

\begin{figure}[!t]
    \centering
    \includegraphics[width=3in]{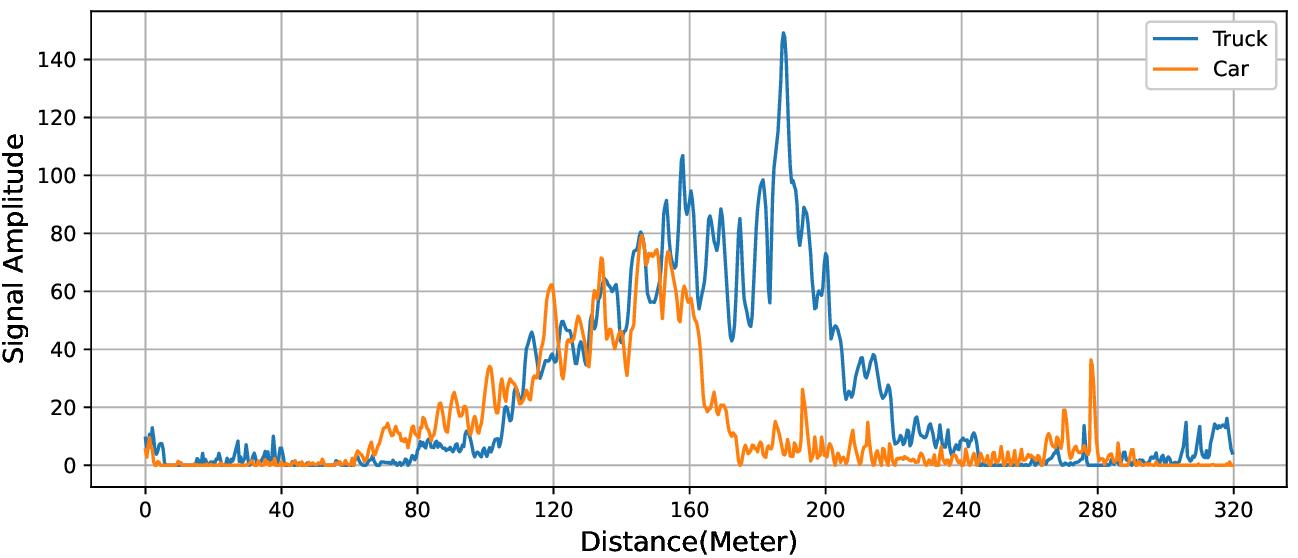}
    \caption{A single vehicle passes through the highway where the optical fiber is laid.}
    \label{single_vehicle}
\end{figure}

\begin{figure}[!t]
    \centering
    \includegraphics[width=3in]{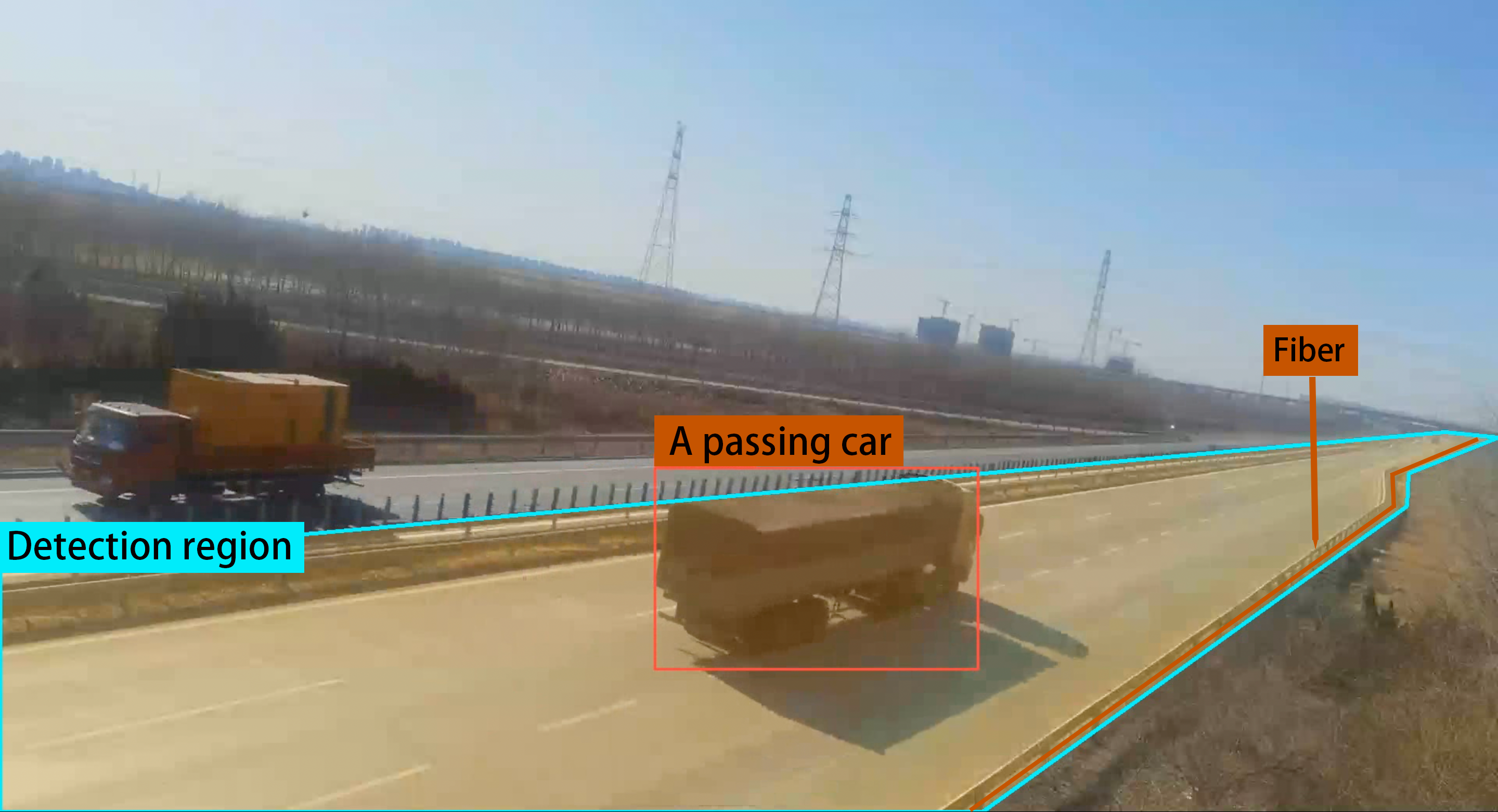}
    \caption{Vehicle information collected by a fixed camera at the end of the road.}
    \label{car_video}
\end{figure}

Our DAS system collected vibration signals of vehicles at a rate of 3 frames per second, with each point at 0.4 meters, for a total duration of 2 hours and 5 minutes, which is equivalent to 7500 seconds. The total length of the optical fiber acquisition distance is 320 meters, corresponding to 800 distance points. When the vehicle traverses the area, the DAS system will record the corresponding vibration signals. We present the vibration signals recorded by the DAS system during the passage of a single truck and a single car through the target area as shown in Fig. \ref{single_vehicle}. At the same time, we recorded corresponding videos at the fixed position of the road, as shown in Fig. \ref{car_video}.

\begin{figure}[!t]
    \centering
    \includegraphics[width=2.5in]{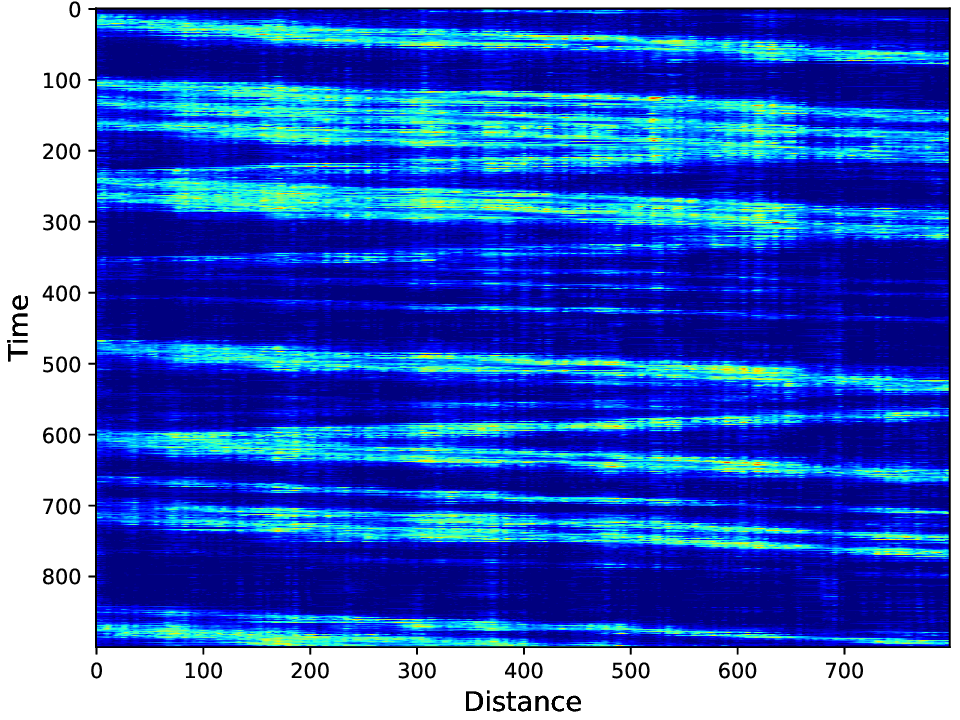}
    \caption{The raw optical fiber waterfall diagram in part of highway scenarios.}
    \label{raw}
\end{figure}

As can be seen from the two types of vehicle driving signals above, their amplitude and duration are quite different. To show the vehicles' trajectory more intuitively, we employ waterfall diagrams to plot the collected signal values, as shown in Fig. \ref{raw}. 

\begin{figure}[t!]
    \centering
    \subfloat[]{
        \includegraphics[width=1.5in]{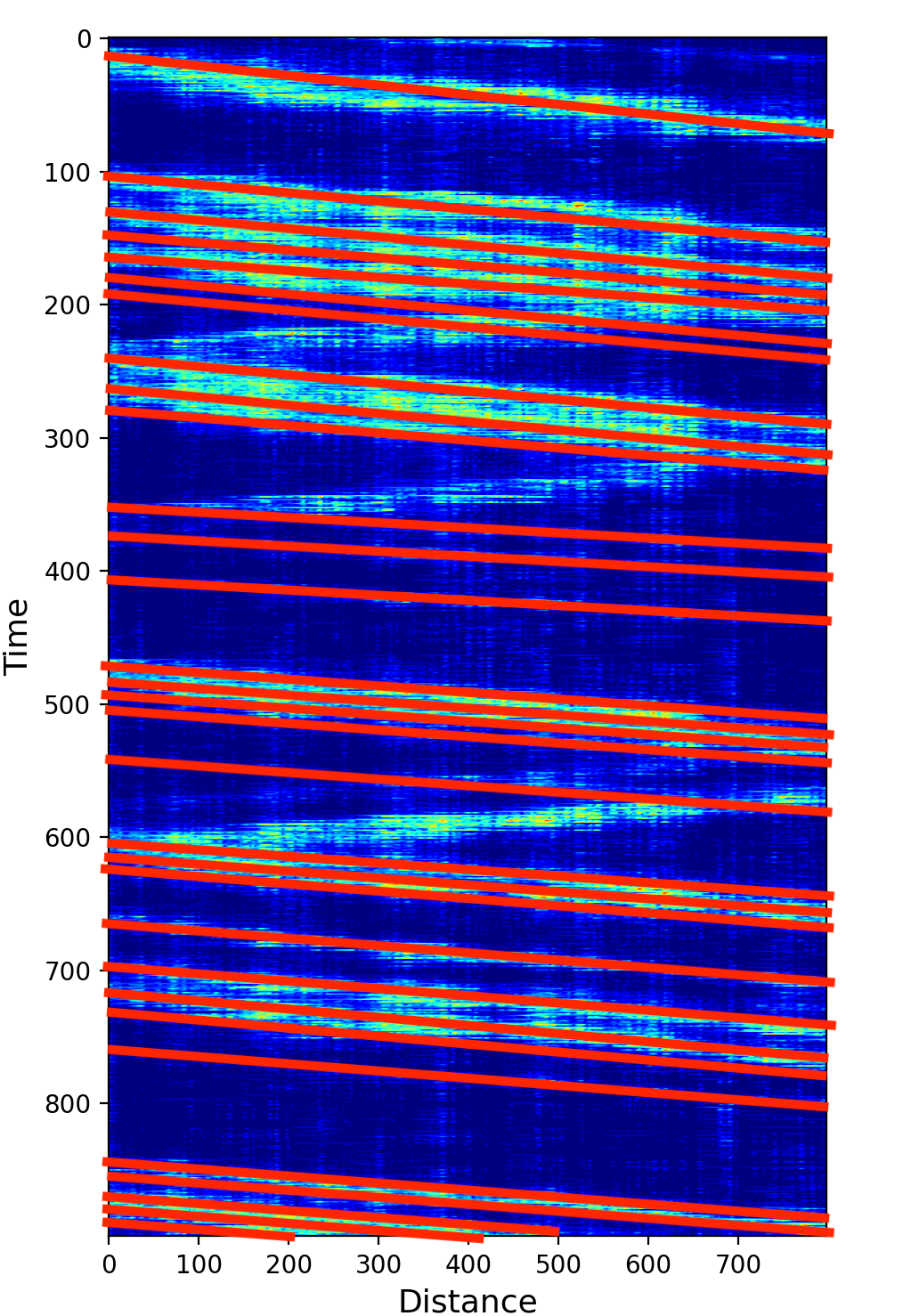}
    }
    \subfloat[]{
        \includegraphics[width=1.5in]{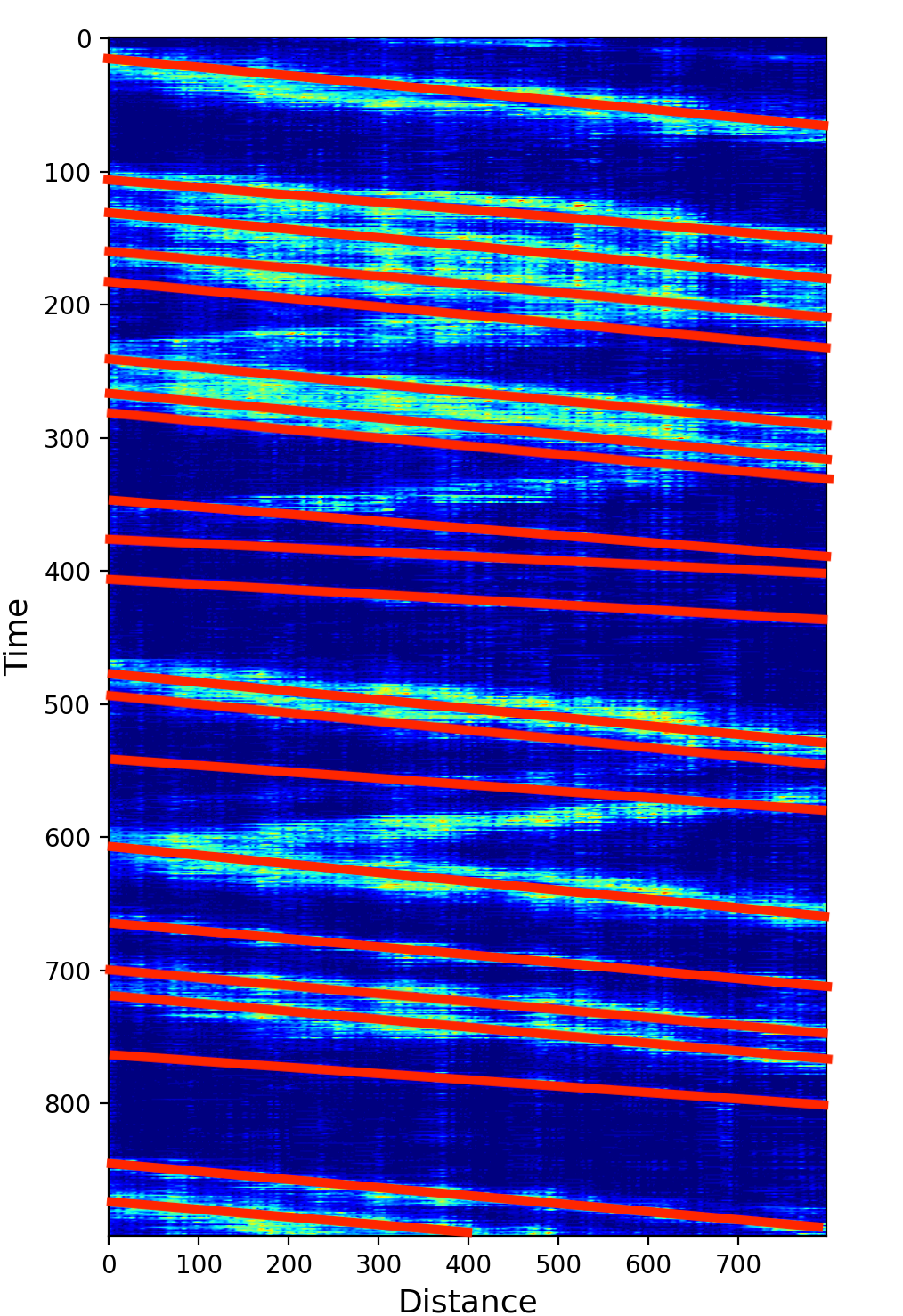}
    }
\caption{A labeled example of vehicle trajectories. (a) real trajectories corresponding to the cameras, (b) fiber trajectories corresponding to human eyes.}
\label{labeled_trajectory}
\end{figure}

In waterfall diagrams, the light blue color represents background noise, whereas the prominently colored lines represent vehicles. The fiber stripes extending from the top left to the bottom right represent vehicles within the target area, while the stripe extending from the top right to the bottom left represents vehicles on the opposite side. Furthermore, the vehicle's trajectory is depicted as an approximately straight line within the diagram, as most vehicles on the highway maintain high and consistent velocities when traveling in lanes with unchanged velocity. The intercept of the straight line represents the starting position of the vehicle, and the absolute value of the slope represents its velocity.

We take Fig. \ref{raw} as an example to calculate the real vehicle flow data corresponding to the video, and plotted 31 vehicle trajectories in Fig. \ref{labeled_trajectory}(a). As can be seen from the diagram, it is obvious that the large vehicle (thicker line) covers the small car (thinner line) when multiple vehicles are parallel, so the trajectory extraction algorithm of the vehicle should first be based on the number of optical fiber stripes visible to human eyes. We counted the number of clearly visible stripes and plotted 21 vehicle trajectories in \ref{labeled_trajectory}(b).

\subsection{Quantitative analysis}

The algorithm performance is validated on real highway scenarios, which are located in Junliang City Test Site, Dongli District, Tianjin, China. In particular, a vehicle trajectory extraction algorithm is adopted to analyze the processed optical fiber signals. We also compared with the Wavelet Filter and Kalman Filter to verify the effectiveness of the proposed online mean estimation algorithm.

\begin{figure}[t!]
    \centering
    \subfloat[$s = 1$]{
        \includegraphics[width=3in]{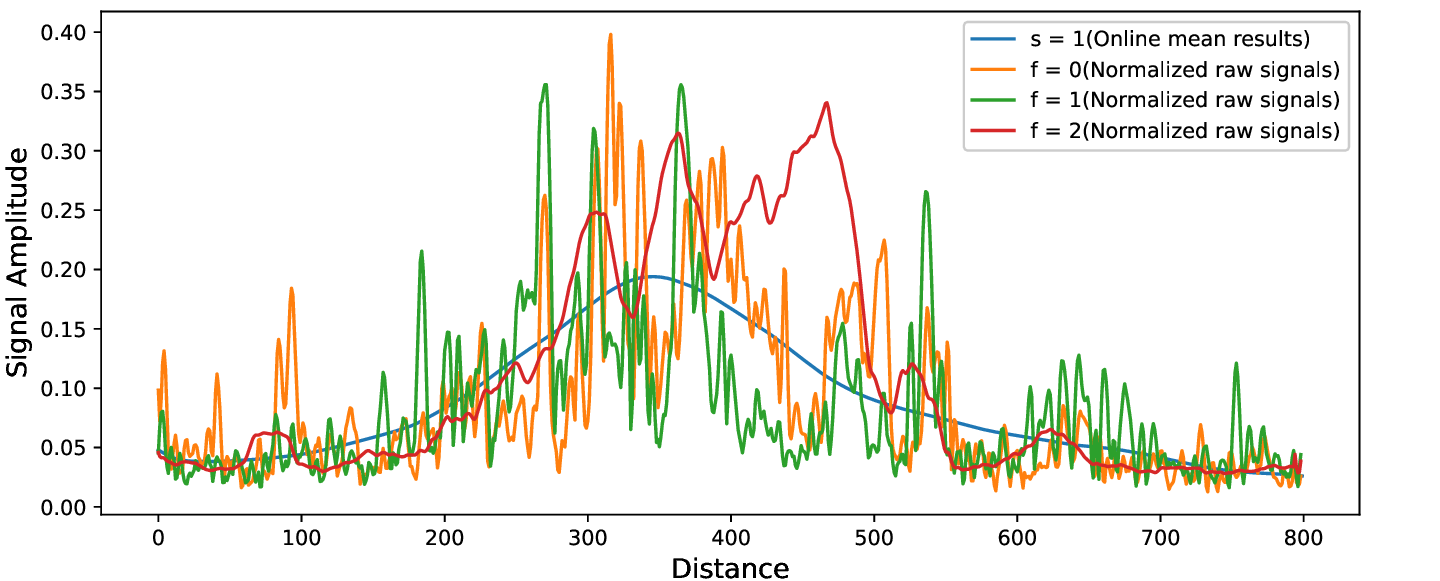}
    }\hfill
    \subfloat[$s = 300$]{
        \includegraphics[width=3in]{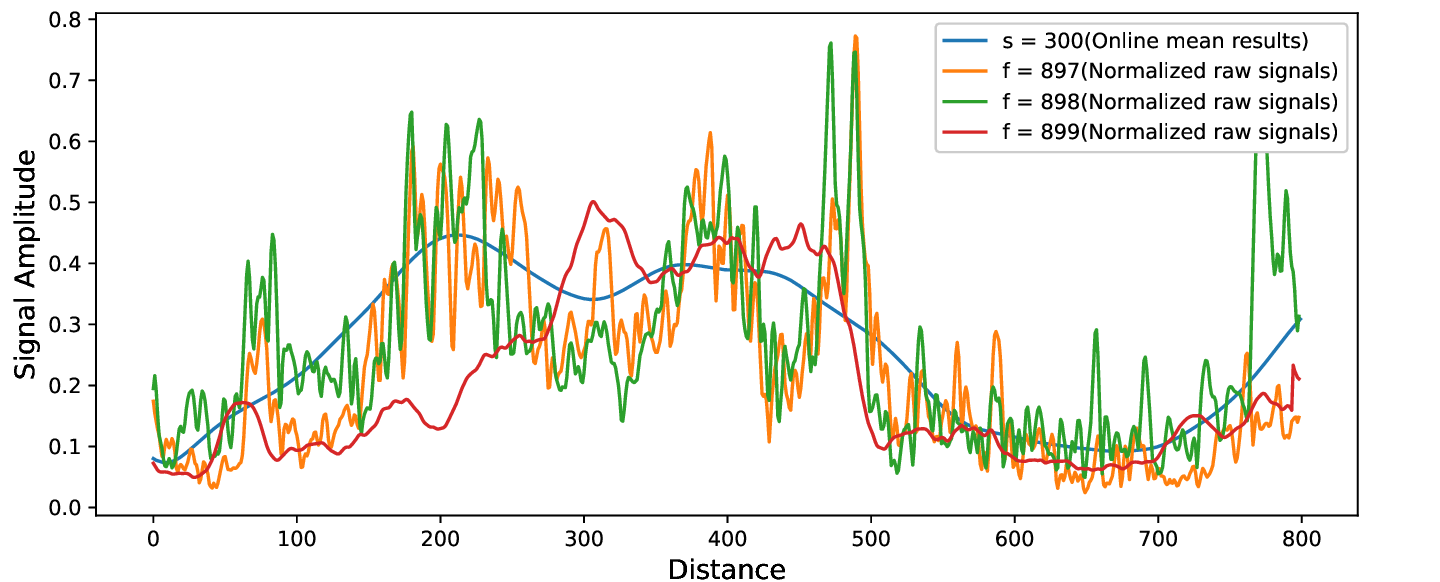}
    }\hfill
\caption{The mean estimation curve of multiple-frame signals compared with normalized raw signals, where $s$ represents the processed seconds and $f$ represents the frame in raw signals.}
\label{mean_curve}
\end{figure}

We randomly selected two seconds of signals for presentation with each second corresponding to three frames, as illustrated in Fig. \ref{mean_curve}. The blue lines represent online mean estimation results for three frames of colored lines. It can be seen that the online mean estimation method can effectively show the trends of multi-frame signals, especially for estimating the peak position of multi-frame data.

Furthermore, we carried out comparative experiments with the filter-based Kalman and Wavelet Filter, employing the same preprocessing steps. Wavelet filtering is a signal processing method based on time-frequency analysis, which can decompose and reconstruct signals at different resolutions. Kalman filtering is a recursive and optimal linear filtering algorithm. Based on Bayesian filtering theory, the Kalman filter is used to estimate the state variables of the system by using the dynamics model and the observation model of the system. Since the filtering process of the two methods cannot directly average the multi-frame signals, the method adopted in this paper is to filter the multi-frame data first and then take the average value for comparison and analysis with online mean estimation.

In real applications, optical fiber signals collected by the DAS system are discretely related to time and distance. Frames Per Second (FPS) are only related to time. On one hand, to process multi-frame data, we present signals as distance-dependent one-dimensional line graphs. On the other hand, to display traffic flow more intuitively, we present signals as three-dimensional waterfall diagrams for vehicle detection algorithms. Here we take two contrasting approaches, considering 1-D signal amplitudes and 2-D waterfall diagram information.

\begin{figure}[htbp]
    \centering
    \subfloat[Raw]{
        \includegraphics[width=3in]{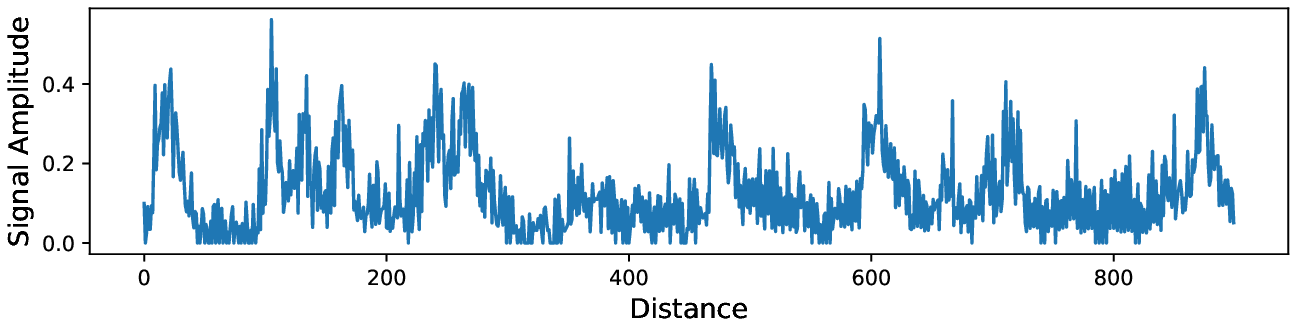}
    }\hfill
    \subfloat[Online]{
        \includegraphics[width=3in]{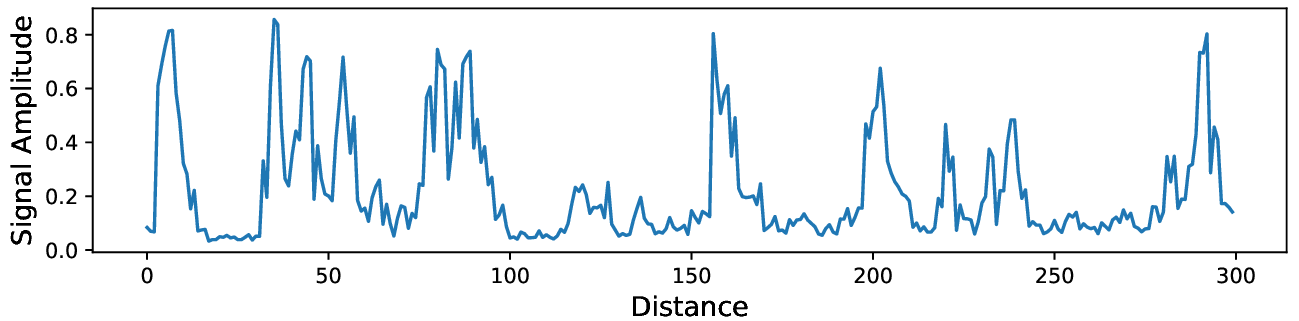}
    }\hfill
    \subfloat[Kalman]{
        \includegraphics[width=3in]{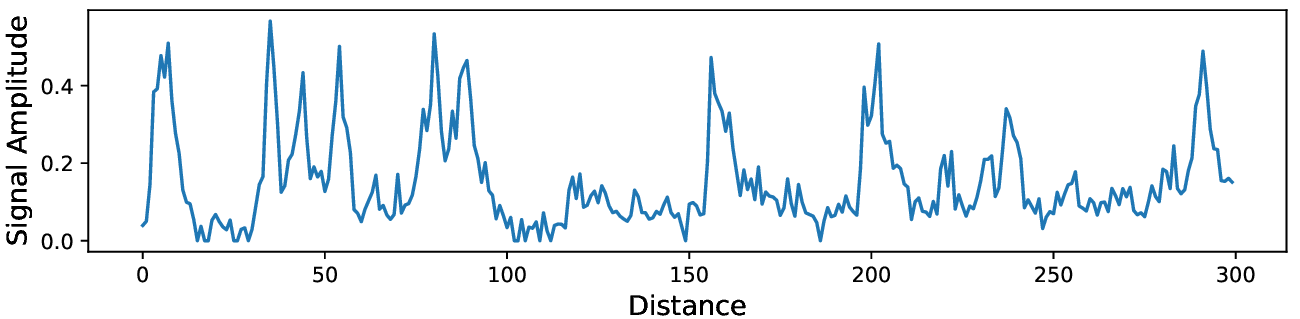}
    }\hfill
    \subfloat[Wavelet]{
        \includegraphics[width=3in]{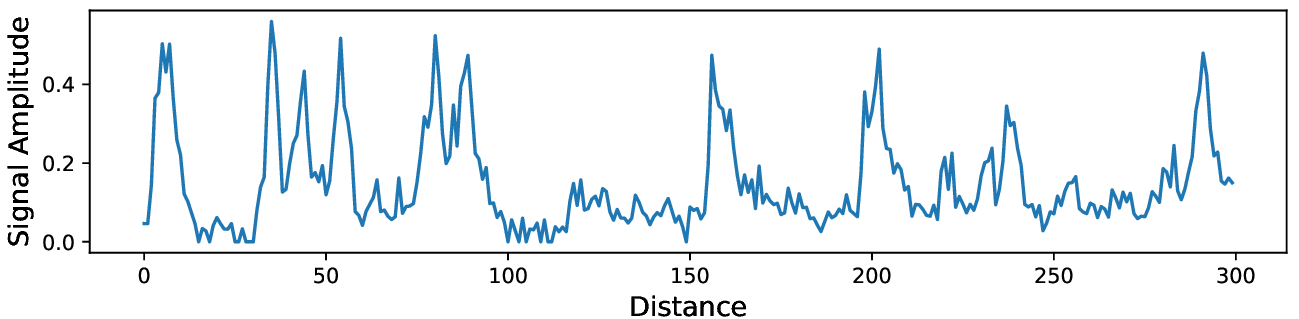}
    }\hfill
\caption{Comparation of 1-D processed signals. (a) Normalized raw signals, (b) Online mean estimation, (c) Kalman Filter, and (d) Wavelet Filter.}
\label{compare_line0}
\end{figure}

In 1-D signal amplitudes analysis, considering the subsequent vehicle trajectory extraction algorithm, we need to determine the vehicle entry situation through the peaks of the starting position. Therefore, we compared the signal processing results of the above algorithm for the first column. As depicted in Fig. \ref{compare_line0}, mean estimation method and the two Filters have excellent noise attenuation effects on the raw signals, and the burr phenomenon of the raw signals is obviously reduced. It is worth mentioning that the algorithm in this paper can amplify the original large truck signals, while the small car signals maintain the original small value, and the bottom noise is also maintained in a relatively stable small base range, so our algorithm shows superior accuracy in extracting vehicle trajectories. 

\begin{figure}[t!]
    \centering
    \subfloat[Raw]{
        \includegraphics[width=1.7in]{pics//compare_2d_raw.eps}
    }
    \subfloat[Online]{
        \includegraphics[width=1.7in]{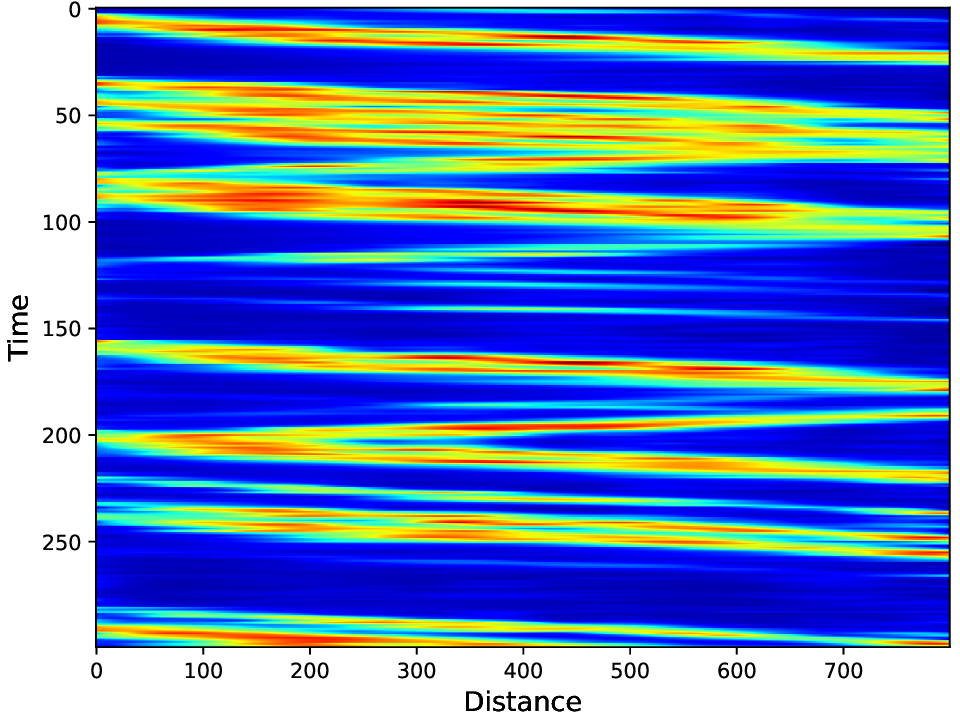}
    }\hfill

    \subfloat[Kalman]{
        \includegraphics[width=1.7in]{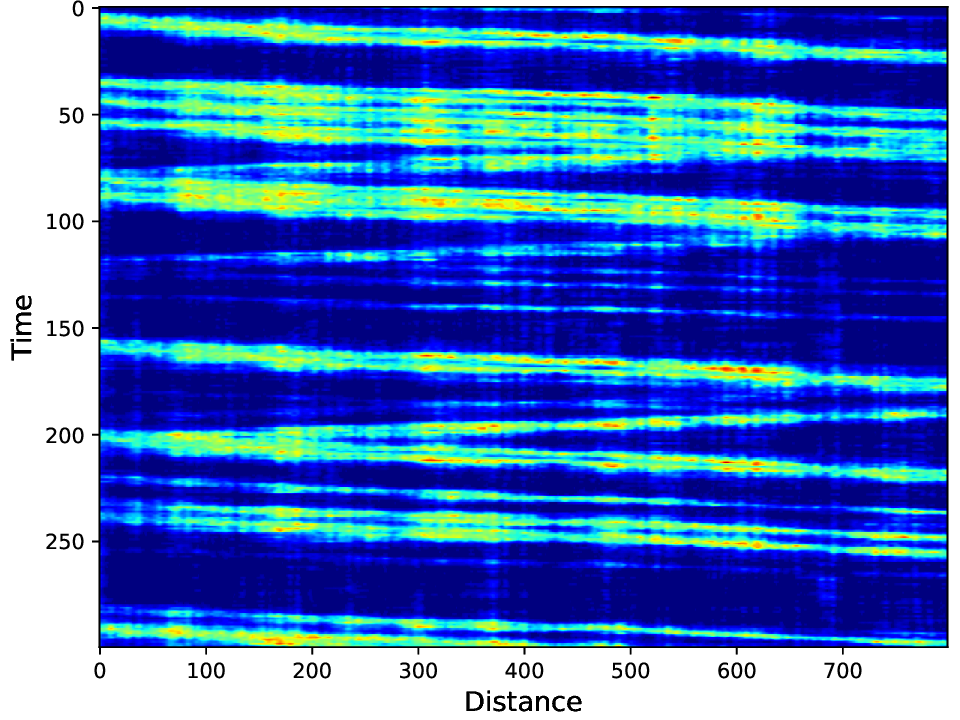}
    }
    \subfloat[Wavelet]{
        \includegraphics[width=1.7in]{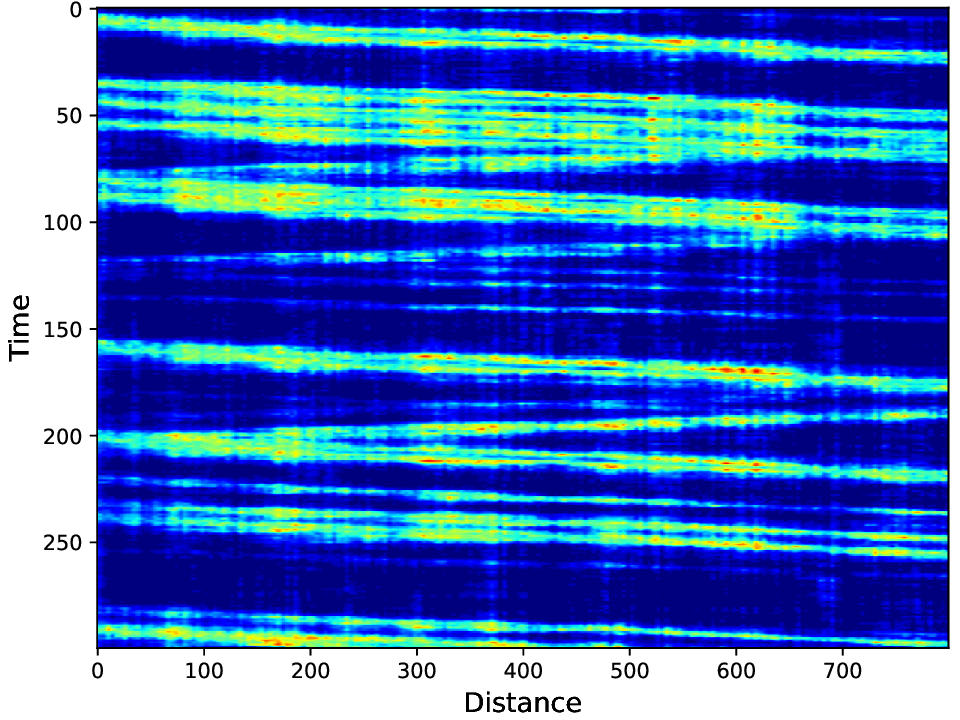}
    }\hfill
\caption{The mean estimation waterfall diagrams of multiple-frame signals compared with normalized raw signals, Kalman Filter, and Wavelet Filter.}
\label{compare_2d}
\end{figure}

In 2-D waterfall diagram information, we can see direct comparisons of different algorithms, as shown in Fig. \ref{compare_2d}. In the fiber waterfall diagrams, the prominently colored lines represent vehicle signals extracted by DAS, and the light blue color represents the background noise. Compared with the other methods, our method attains smoother diagrams, with obvious background noise removal effects and stronger signals for the vehicles.


\begin{figure}[t!]
    \centering
    \subfloat[Raw]{
        \includegraphics[width=1.7in]{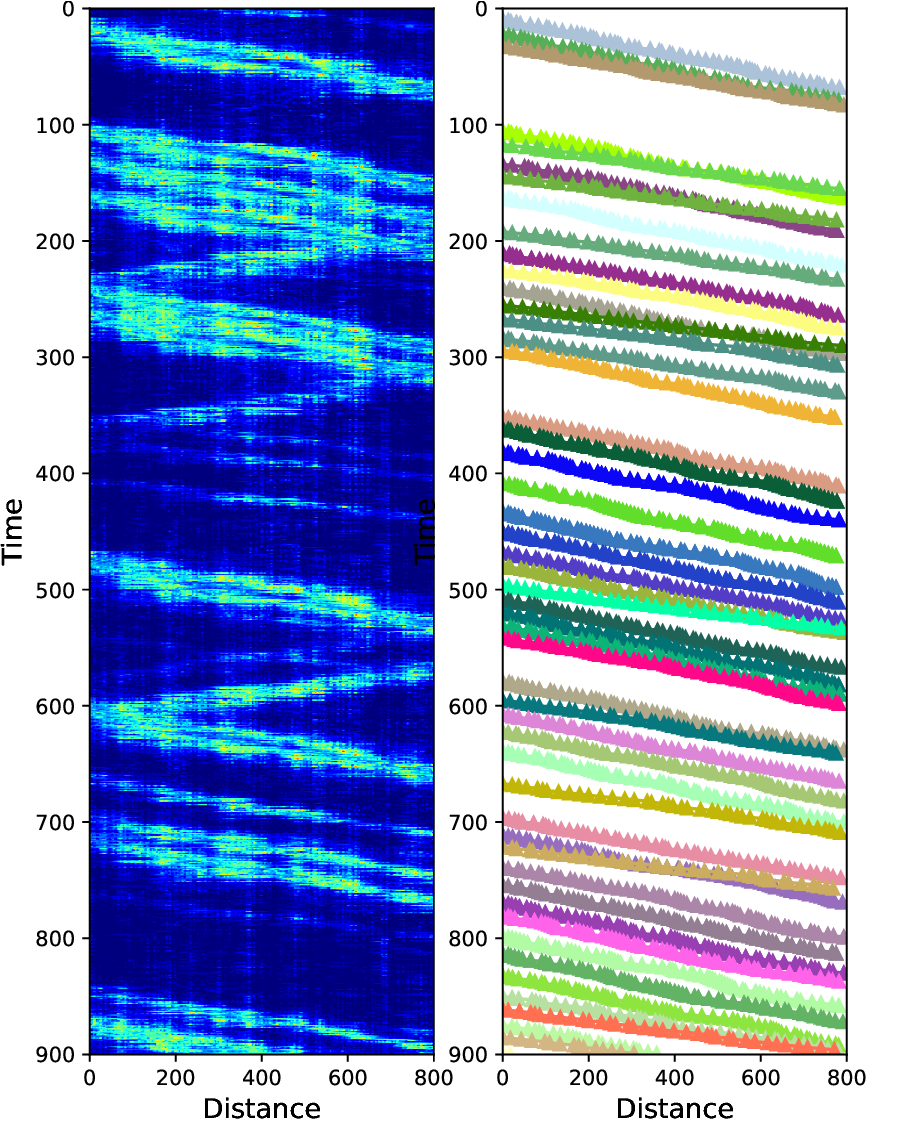}
    }
    \subfloat[Online]{
        \includegraphics[width=1.7in]{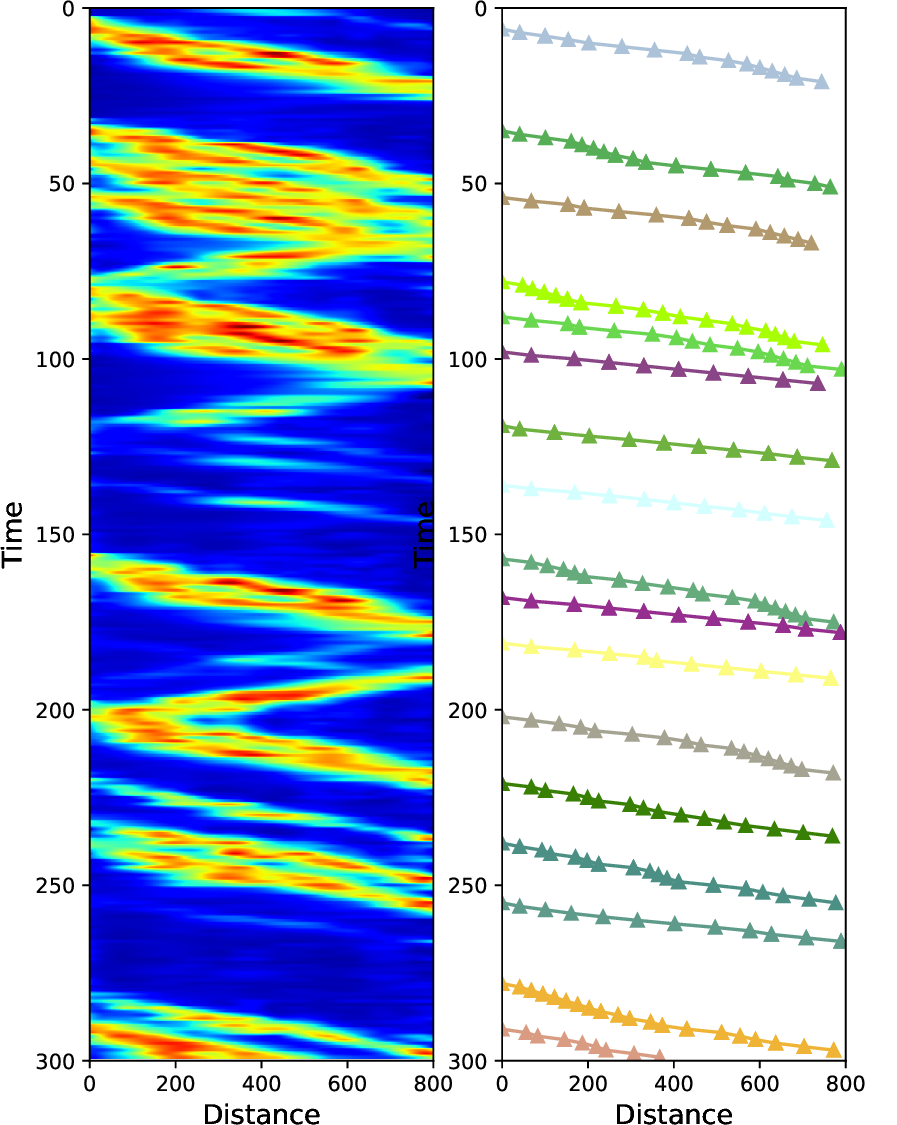}
    }\hfill

    \subfloat[Kalman]{
        \includegraphics[width=1.7in]{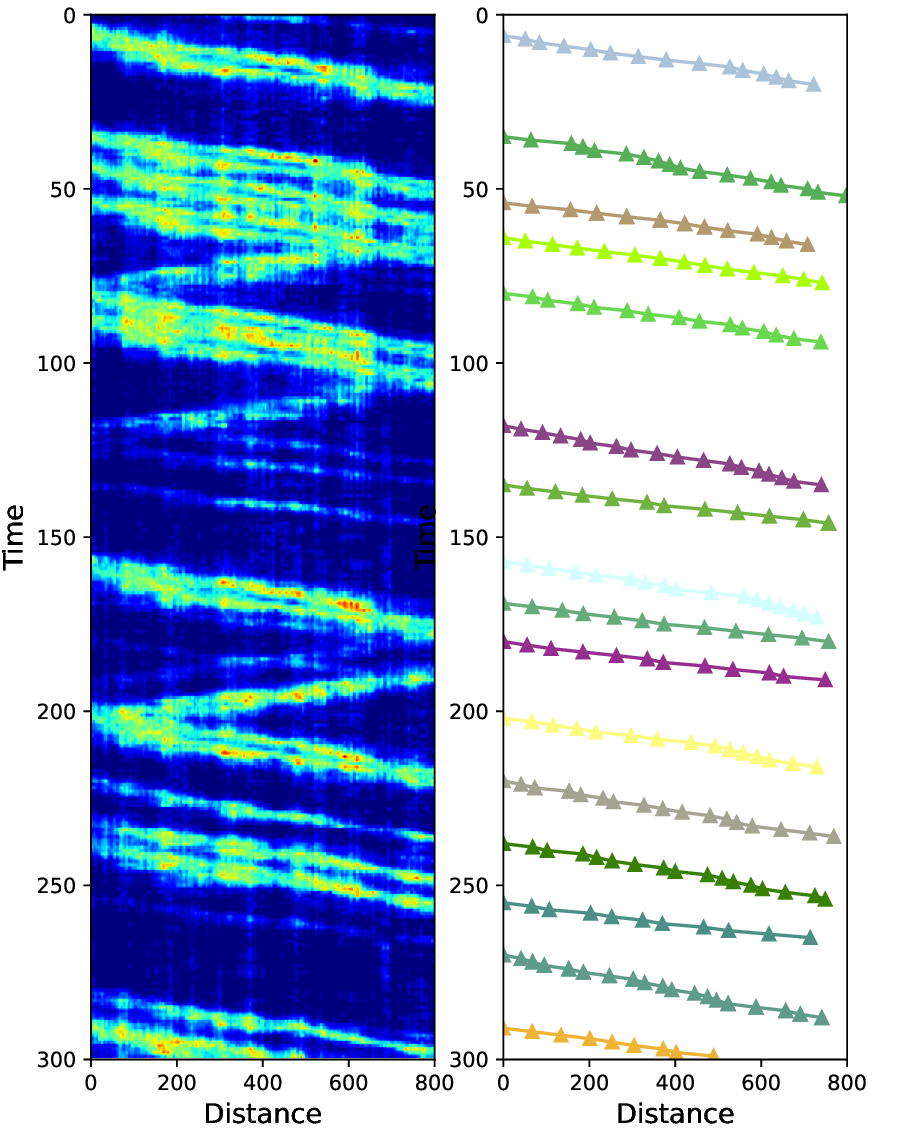}
    }
    \subfloat[Wavelet]{
        \includegraphics[width=1.7in]{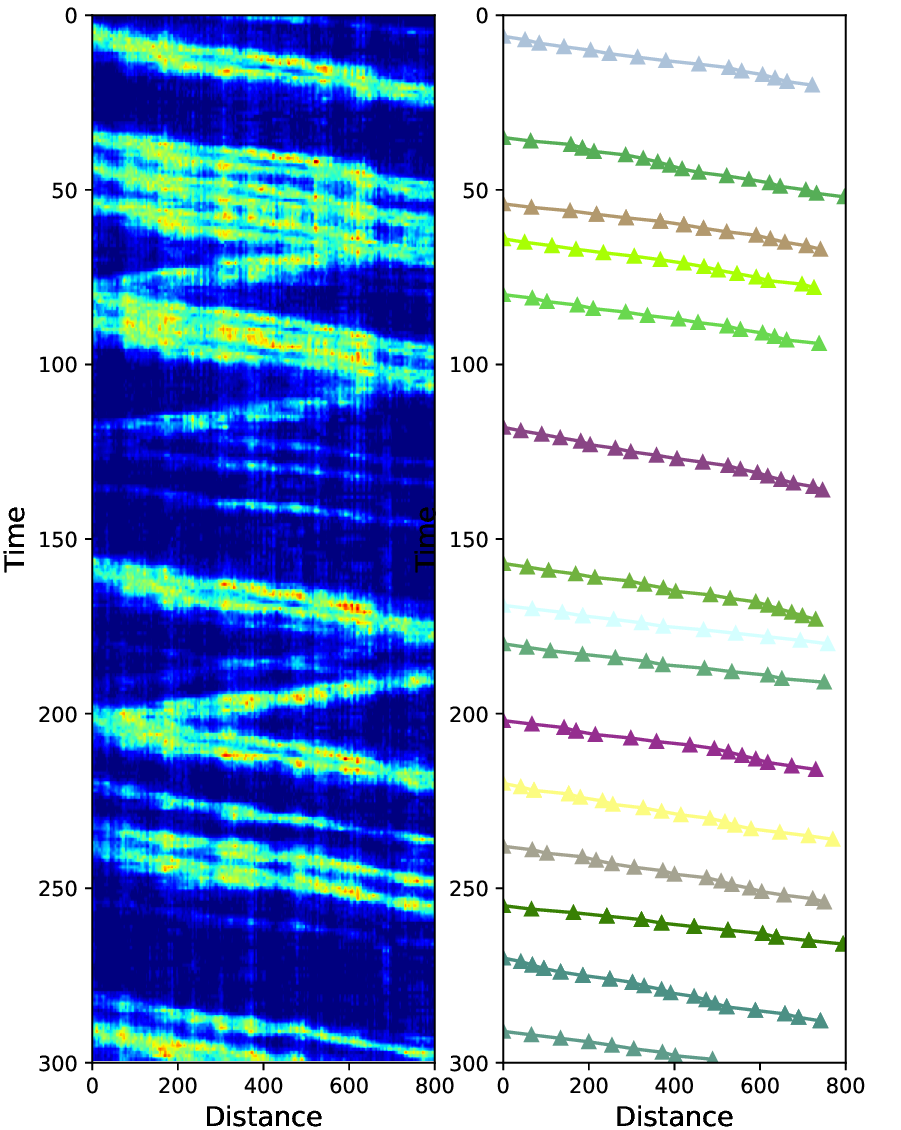}
    }\hfill
\caption{The vehicle trajectory extraction results of multiple-frame signals compared with normalized raw signals, Kalman Filter, and Wavelet Filter.}
\label{compare_trajectory}
\end{figure}

Based on preprocessed multi-frame signals, we also utilize a vehicle trajectory extraction algorithm with the same parameters to examine the effectiveness of subsequent structured data extraction. As displayed in Fig. \ref{compare_trajectory}, compared to normalized raw signals, the online estimation method and the two filter-based methods are all able to extract trajectories consistent with fiber stripes. In particular, the online mean estimation algorithm is more precise in extracting finer fiber stripes.
 
To further quantify the effectiveness and superiority of online preprocessing results, the algorithm is evaluated by real signals by the DAS system. We randomly selected 5min of signals for comparative experiments. Through the corresponding recordings of the camera, a total of 31 vehicles (including 11 trucks and 20 cars) passes by, but because the vehicle is too close or the truck's signal overcovers the small car, etc., we give a true labeled trajectory of 21 vehicles as shown in \ref{labeled_trajectory}(b). As can be seen in Table. \ref{trajectory_statistics}, although all vehicle tracks are not entirely extracted by the three methods, the online algorithm achieves the highest accuracy in terms of low missing rate and error rate. It can be proved that our online denoising algorithm has a strong application value in the subsequent feature extraction algorithm.

\begin{table}[!t]
    \caption{Trajectory statistics of different preprocessing methods.}
    \centering
    \renewcommand\arraystretch{1.2}
    \begin{tabular}{cccccc}
    \hline
     & Total count & Correct  & Missing & Wrong & Accuracy\\
    \hline
    Online &17 & 17 & 4 & 0 & 80.95\% \\
    Kalman &18 & 15 & 6 & 3 & 71.43\% \\
    wavelet & 15& 13 & 8 & 2 & 61.90\% \\
    \hline
    \label{trajectory_statistics}
    \end{tabular}
\end{table}

\section{Conclusion}
\label{conclusion}

In this paper, we have demonstrated the application of optical fiber-based Distributed Acoustic Sensors (DAS) systems to monitor traffic flow in real highway scenarios. An online mean estimation method is utilized to deal with multi-frame problems in signal preprocessing. Compared with existing filter-based methods, such as Kalman Filter and Wavelet Filter, the online method achieves real-time processing while saving computing memory, demonstrating its superiority. The major conclusions and areas for improvement in this research are as follows:

(1) The problem of multi-frame signals is characterized by a non-parametric regression model, and the mean function is fitted by a Local Polynomial Regression rather than averaging directly, which cannot be solved by filter-based methods. 

(2) The online method using only summary statistics of previous data, instead of historical raw data, is of critical importance in handling big data as far as computing memory and speed are concerned.

(3) In this paper, signals are stored as summary statistics and applied to mean estimation problems. In addition, other applications of statistics in the field of transportation can be explored, such as spatio-temporal prediction, outlier detection, etc.
\section*{Declaration of Competing Interest}
The authors declare that they have no known competing financial interests or personal relationships that could have appeared to influence the work reported in this paper.

\section*{Acknowledgments}

This research was funded by Tianjin Yunhong Technology Development (Grant number: 2021020531).

\bibliographystyle{ieeetr}
\bibliography{reference}
\end{document}